\pdfoutput=1

\documentclass[fleqn,usenatbib,usedcolumn,letters]{mnras}
\usepackage{natbib}

   \usepackage[british]{babel}             
   \usepackage{newtxtext}                  
   \usepackage[slantedGreek]{newtxmath}    
  
   \let\la=\lesssim 
  
   \usepackage[T1]{fontenc}               
   \usepackage{graphicx}

\newcommand{\mpc}{\,\mathrm{Mpc}\,h^{-1}}

\title{On the [$\alpha$/Fe]--[Fe/H] relations in early-type galaxies}
\author[F. Vincenzo, C. Kobayashi \& P. Taylor]{Fiorenzo Vincenzo$^{1}$\thanks{f.vincenzo@herts.ac.uk}, 
Chiaki Kobayashi$^{1}$ \& Philip Taylor$^{2,3}$
\\
$^{1}$Centre for Astrophysics Research, University of Hertfordshire, College Lane, Hatfield, AL10 9AB, UK \\
$^{2}$Research School of Astronomy and Astrophysics, Australian National University, Canberra, ACT 2611, Australia \\
$^{3}$ARC Centre of Excellence for All Sky Astrophysics in 3 Dimensions (ASTRO 3D), Australia}

\begin{document}

\date{Accepted 2018 July 7. Received 2018 June 22; in original form 2018 April 27}

\pagerange{\pageref{firstpage}--\pageref{lastpage}} \pubyear{2018}

\maketitle

\label{firstpage}


\begin{abstract}

\noindent We study how the predicted $[\alpha/\text{Fe}]$--$[\text{Fe/H}]$ relations in early-type galaxies vary as functions of their stellar masses, 
ages and stellar velocity dispersions, 
by making use of cosmological chemodynamical simulations with feedback from active galactic nuclei. 
Our model includes a detailed treatment for the chemical enrichment from dying stars, core-collapse supernovae (both Type II and hypernovae) and 
Type Ia supernovae. 
At redshift $z=0$, we create a catalogue of $526$ galaxies, among which we determine $80$ early-type galaxies. 
From the analysis of our simulations, we find $[\alpha/\text{Fe}]$--$[\text{Fe/H}]$ relations similar to the Galactic bulge. We also find that, 
in the oldest galaxies, Type Ia supernovae start to contribute at higher $[\text{Fe/H}]$ than in the youngest ones. 
On the average, early-type galaxies with larger stellar masses (and, equivalently, 
higher stellar velocity dispersions) have 
higher $[\alpha/\text{Fe}]$ ratios, at fixed $[\text{Fe/H}]$.
This is qualitatively consistent with the recent observations of Sybilska et al., but quantitatively there are mismatches, which might require 
stronger feedback, sub-classes of Type Ia Supernovae, or a variable initial mass function to address. 
\end{abstract}




\section{Introduction} \label{sec:intro}
The observed $[\alpha/\text{Fe}]$ ratio historically represents one of the most used chemical diagnostics 
for Galactic and extragalactic astro-archaeology studies (e.g., \citealt{kobayashi2016}).
The $[\alpha/\text{Fe}]$--$[\text{Fe/H}]$ chemical abundance pattern was initially shown for the solar neighborhood 
and has been used to understand the origin of Type Ia Supernovae (SNe Ia) (e.g., \citealt{matteucci1986,kobayashi1998}). 
Later, the observed $[\alpha/\text{Fe}]$ ratios in nearby elliptical galaxies were used to constrain their formation scenario \citep{matteucci1994,thomas2010}. 
Although various types of models have been suggested 
\citep{pipino2004,kobayashi2004,delucia2006,taylor2015a,delucia2017,demasi2018}, 
the observed correlations between $[\alpha/\text{Fe}]$ 
and other physical and chemical properties of early-type galaxies (ETGs) still represent unsolved problems for galaxy formation and evolution in a cosmological context.  

In our Milky Way (MW), the $[\alpha/\text{Fe}]$--$[\text{Fe/H}]$ abundance ratio pattern (which can be separately measured 
in the halo, bulge, thick and thin disc stellar components) can be effectively used to probe the characteristic star formation histories (SFHs) of the 
different MW stellar components \citep[e.g.,][]{matteucci2012}. 
The specific trend of 
$[\alpha/\text{Fe}]$ versus $[\text{Fe/H}]$ in our Galaxy \citep{kobayashi2011b} and the similar trends in 
its dwarf spheroidal galaxy satellites \citep{lanfranchi2004} 
can be explained as the consequence of the different channels for the nucleosynthesis 
of $\alpha$-elements (O, Mg, Ne, Si, S, and Ca) and iron.
On the one hand, $\alpha$-elements are mainly produced on short timescales ($\sim10^{6}\,\text{yr}$) 
by core-collapse supernovae (SNe), which 
can also eject some iron into the interstellar medium (ISM) of galaxies; in particular, core-collapse SNe represent 
the only source of Fe in the earliest stages of the galaxy evolution. 
On the other hand, the bulk of iron is produced by SNe Ia on longer 
timescales, causing $[\alpha/\text{Fe}]$ to decrease.
SN Ia progenitors are not well understood yet, and 
there may be multiple channels, but -- in any case -- the timescale is at least $\sim  0.01\,\text{Gyr}$ \citep[e.g.,][]{kobayashi2009}. 

In elliptical galaxies, 
$[\alpha/\text{Fe}]$ and $[\text{Fe/H}]$ can be estimated from absorption lines 
\citep[e.g.,][]{faber1973,worthey1994,thomas2003b,cervantes2009},
and the correlations with a variety of different galaxy properties 
such as stellar mass, age, and environments have been shown \citep[e.g.,][]{thomas2010,kuntschner2010,spo10}.
Recently, \citet{kriek2016} 
showed very high $[\alpha/\text{Fe}]$ ratios in ETGs at redshift $z=2$; this observation made it harder to explain the formation of massive galaxies in a cosmological context. 
Moreover, with  integral-field unit (IFU) observations of nearby ETGs, \citet{sybilska2018} showed a mass-dependence of the $[\text{Mg/Fe}]$--$[\text{Fe/H}]$ relation.
In this Letter, using our cosmological chemodynamical simulations that include feedback from active galactic nuclei (AGN) and a detailed treatment for 
the chemical enrichment, 
we aim to understand the observational result in \citet{sybilska2018}.

This Letter is structured as follows. In Section \ref{sec:model}, we describe our cosmological chemodynamical simulation and the methods of analysis. 
In Section \ref{sec:results}, 
we present our results, mostly focusing on how the predicted $[\alpha/\text{Fe}]$--$[\text{Fe/H}]$ relations in ETGs 
vary as functions of their stellar mass and age. 
In Section \ref{sec:conclusions}, we draw our conclusions. 

\begin{figure}
\centering
\includegraphics[scale=0.5]{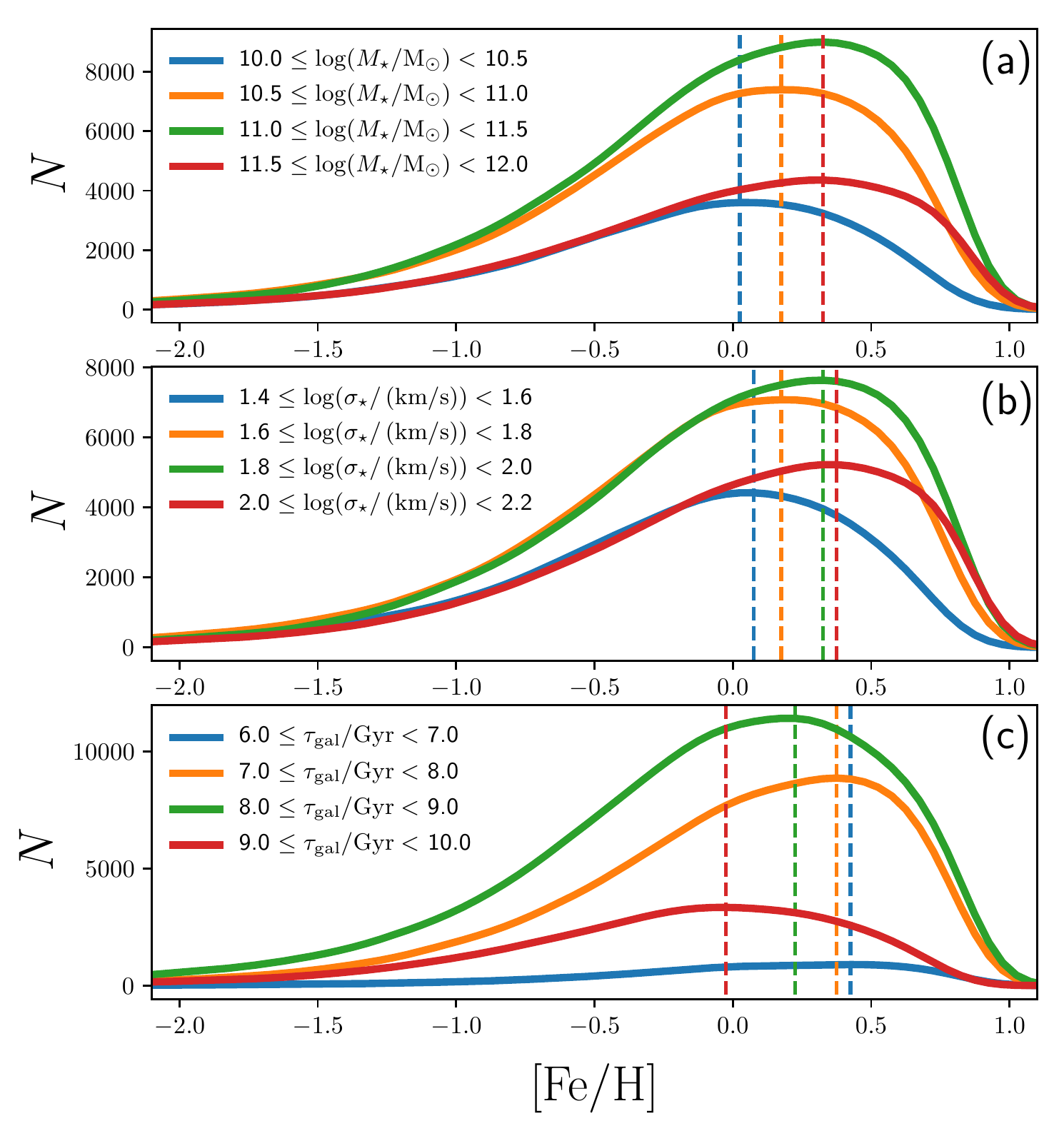}
\caption{The $[\text{Fe/H}]$ distribution function of all the stellar populations in our simulated ETGs within a given	stellar mass bin (panel a), stellar velocity 
dispersion bin (panel b) and galaxy age bin (panel c). The vertical dashed lines correspond to the $[\text{Fe/H}]$ abundances at the maximum of each distribution.   }
\label{fig:hist}
\end{figure}

\section{Model, simulations and methods} \label{sec:model}

\subsection{The model}
We analyse two cosmological simulations run by our chemodynamical code based on \textsc{Gadget-3}, 
a smoothed-particle hydrodynamics code \citep{springel2005}. 
Our code has been developed by \citet{kobayashi2004,kobayashi2007,taylor2014} to 
include the relevant baryon physical processes such as 
UV background heating; metal-dependent radiative cooling; star formation activity 
with the corresponding thermal energetic feedback from 
stellar winds and SN explosions; formation of black hole (BH) seeds as the debris of the first stars; the growth and feedback from AGN; 
finally, chemical enrichment from AGB stars and core-collapse SNe and SNe Ia. 
The thermal energetic feedback and the nucleosynthetic products are distributed, kernel-weighted,  
to a fixed number of neighbour gas particles, $N_{\text{FB}}=72$. 

We adopt the mass and metallicity-dependent stellar yields of \citet{kobayashi2011b} and the metallicity-dependent stellar lifetimes 
as in \citet{kobayashi2007}. Finally, we assume the universal \citet{kroupa2008} initial mass function (IMF), defined in the mass range $0.01\le m \le 120\,\text{M}_{\sun}$.

\textit{SN Ia model} --- 
For the predictions of [$\alpha$/Fe] ratios, the most important factor is the SN Ia model. Our model is based on Chandrasekhar mass explosions of 
C+O white dwarfs (WDs) in single-degenerate systems. 
Because of the WD winds, our SN Ia rate depends on the metallicity, and becomes zero below the progenitor $\text{[Fe/H]} =-1.1$ \citep{kobayashi1998},
This model gives an excellent agreement with the observed [$\alpha$/Fe] and [Mn/Fe] in the solar neghbourhood.
The shortest lifetime is $0.5\,\text{Gyr}$ at $\text{[Fe/H]}=-1.0$, and becomes $0.1\,\text{Gyr}$ at $\text{[Fe/H]}=0.0$ \citep[see][for the details]{kobayashi2009}.
In nearby galaxies, the lifetime distribution is very similar to the observed delay-time distribution of nearby SNe Ia, and also to that of a simple model of double degenerate systems \citep{mao14}.
Note that sub-classes of SNe Ia \citep{kobayashi2015} are not included (see also Section \ref{subsec:comparison}). 

\textit{AGN feedback} --- 
In order to model massive ETGs, it is necessary to include additional feedback, and the feedback from AGN is the most likely 
as it can work in a deep potential well. The co-evolution between AGN and galaxy stellar components is supported from the observed BH mass--bulge mass relation \citep[e.g.,][]{mag98}.
Our AGN model successfully reproduces a number of observed properties of galaxies, 
including the cosmic SFRs \citep{taylor2014}, galaxy size-mass relations, mass-metallicity relations (MZR; \citealt{taylor2015a,taylor2016}), 
and metallicity radial gradients \citep{taylor2017}. 
Here we briefly recall that BHs are assumed to form from metal-free gas with volume mass 
density above $\rho_{c} = 0.1\,h^{2}\,m_{\text{H}}\,\text{cm}^{-3}$. 
The initial seed mass of the primordial BHs is $M_{\text{BH,seed}}=1000\,h^{-1}\,\text{M}_{\sun}$; then, the BH mass 
can grow as a function of time via Eddington-limited, Bondi-Hoyle gas accretion. 
The mass growth of the BHs is due to gas accretion from the surrounding ISM and mergers with other BHs, which -- in turn -- 
trigger the AGN activity. The AGN rate of thermal energy feedback is directly proportional to the accretion rate onto the BH.

\subsection{The simulations} \label{sec:simulations}

We use two simulations of a cubic volume for the standard $\Lambda$-cold dark matter Universe, with periodic boundary conditions and 
cosmological parameters from the nine-year Wilkinson Microwave Anisotropy Probe \citep[$\Omega_0=0.28$, $\Omega_{\Lambda}=0.72$, $\Omega_{\mathrm{b}}=0.046$,
$H_0 =100\times h = 70$\,km\,s$^{-1}$\,Mpc, and $\sigma_8 = 0.82$]{hinshaw2013}.

Simulation A is the same simulation as used in 
\citet{taylor2016}, with an initial condition giving rise to a strong central concentration of galaxies at redshift $z=0$.  
This simulation has a volume of ($25\,\mpc)^3$ and total number of dark matter (DM) and gas particles 
$N_{\text{DM}} = N_{\text{gas}} = (240)^{3}$. 
The mass resolutions 
are $M_{\text{gas}}=1.44 \times 10^{7}\,h^{-1}\,\text{M}_{\sun}$ and 
$M_{\text{DM}}=7.34 \times 10^{7}\,h^{-1}\,\text{M}_{\sun}$ for the gas and DM particles, respectively; the gravitational softening length of gas is $\epsilon_{g}=1.125\,\text{kpc}\,h^{-1}$. 

Simulation B corresponds to the {\ttfamily 010mpc128} simulation in \citet{taylor2014}, but with the same initial condition as in \citet{kobayashi2007}.
This shows a weak concentration of galaxies in the box at redshift $z=0$, which corresponds to the field environment. This simulation 
has a volume of $(10\,\mpc)^3$ and $N_{\text{DM}} = N_{\text{gas}} = (128)^{3}$. 
The mass resolutions are 
$M_{\text{gas}} = 6.09 \times 10^{6}\,h^{-1}\,\text{M}_{\sun}$ and 
$M_{\text{DM}} = 3.10 \times 10^{7}\,h^{-1}\,\text{M}_{\sun}$, with $\epsilon_{g}=0.844\,\text{kpc}\,h^{-1}$. 

Each star particle in the simulation corresponds to a simple stellar population 
with fixed age and metallicity; the typical masses of the star particles are in the range between $3$-$7\times10^{6}\,h^{-1}\,\text{M}_{\sun}$ 
(see \citealt{kobayashi2007} for the detailed model of star formation). 
In this Letter, the star particles within a given simulated galaxy are called  ``stellar populations''.

\begin{figure}
\centering
\includegraphics[scale=0.5]{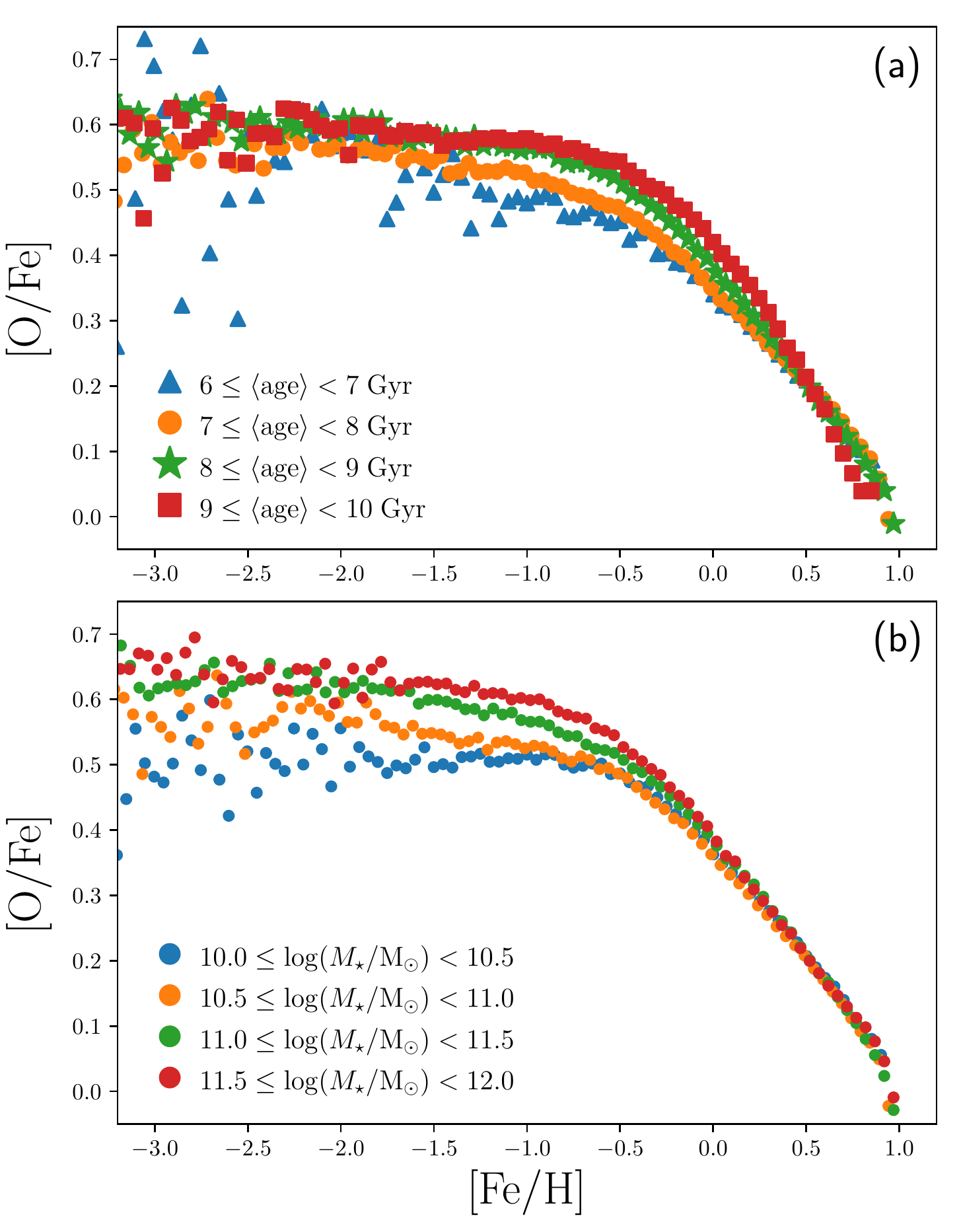}
\caption{The $[\text{O/Fe}]$--$[\text{Fe/H}]$ chemical evolution tracks of all our simulated ETGs as functions of the average galaxy stellar age (panel a) and 
total galaxy stellar mass (panel b).  }
\label{fig:ofe_feh}
\end{figure}

\subsection{Methods} \label{sec:methods}
We adopt the code \textsc{Rockstar} \citep{behroozi2013}, 
to create a catalogue of DM haloes from simulations A and B at redshift $z=0$. 
Due to the limited 
resolutions of our simulations, we only include in the catalogue the DM haloes with virial 
masses $M_{\text{vir,DM}}\ge10^{11}\,\text{M}_{\sun}$.

We isolate and analyse $486$ galaxies and $40$ galaxies in the simulations A and B, respectively.
In total, we have $526$ galaxies in our catalogue,
with total stellar mass $9.0 \lesssim \log(M_{\star}/\text{M}_{\sun}) \lesssim 11.8$. This covers the mass range of the observations. 

In the catalogue, each galaxy is defined by considering all the gas and star particles within the scale radius of a Navarro-Frenk-White profile fitted to the dark matter halo. 
For each galaxy, we visually check that our automated algorithm properly 
works in selecting and isolating all the star and gas particles within the scale radius of the host DM halo, in order to exclude any external contributions 
from close satellites or merging companions. 

ETGs are determined from the same criterion of \citet{fossati2015} (see also \citealt{spitoni2017} and \citealt{taylor2017}); 
namely, one galaxy is defined as ``passive'' if its specific star formation rate (SFR) 
$\text{sSFR}=\text{SFR}/M_{\star} < b / t_{z0} \approx 2.17 \times 10^{-11}\,\text{yr}^{-1}$, where 
$b= \text{SFR}/ \langle \text{SFR} \rangle = 0.3$ is the so-called birthrate parameter \citep{sandage1986,franx2008} 
and $t_{z0}$ is the age of the Universe at redshift $z=0$. Eventually we have $80$ ETGs in our catalogue, $61$ from simulation A and 
$19$ from simulation B. There is no significant difference in the SFRs and MZR between simulations A and B \citep{kacprzak2015}, although 
massive galaxies are found only in simulation A.

In order to compare with the observed mass-dependence 
of the $[\text{Mg/Fe}]$--$[\text{Fe/H}]$ relations in ETGs,
we applied the following scheme of our analysis, similar to that in \citet{sybilska2018}.

\begin{enumerate}  

\item For each ETG in the catalogue, we compute $[\text{Fe/H}]$ and $[\text{O/Fe}]$ 
of all its stellar populations. We assume 
the \citet{asplund2009} solar abundances. 

\item The ETGs are divided into bins of either $\log(M_{\star}/\text{M}_{\sun})$, average stellar age, 
or average stellar velocity dispersion. We use bins of width $\Delta_{\log(M_{\star}/\text{M}_{\sun})} = 0.5$, 
$\Delta_{\tau} = 1\;\text{Gyr}$, and $\Delta_{\log(\sigma/(\text{km}\,\text{s}^{-1}))} = 0.2$, respectively.

\item For each bin of $\log(M_{\star}/\text{M}_{\sun})$, age or $\log(\sigma/(\text{km}\,\text{s}^{-1}))$, all the galaxy stellar populations of (1) 
are binned according to their $[\text{Fe/H}]$ abundance, by assuming a bin width 
$\Delta_{[\text{Fe/H}]} = 0.05$; within each $[\text{Fe/H}]$  bin, we compute the average $[\text{O/Fe}]$ ratio. There is no significant change if we weight by mass or luminosity.  

\end{enumerate}

\section{Results} \label{sec:results}

In Figure \ref{fig:hist}(a), we show how the iron abundance distribution function of all the stellar populations in our simulated ETGs is predicted 
to vary as a function of the total galaxy stellar mass; different colours in the 
figure correspond to galaxies lying in different bins of $\log(M_{\star}/\text{M}_{\sun})$. 
We find that ETGs with increasing stellar mass have the distribution peaked towards higher and higher [Fe/H]. 

Since there is a straight correlation between the total galaxy stellar mass and the stellar velocity dispersion, $\sigma$, we also find that galaxies 
with larger values of $\sigma$ have stellar populations with higher [Fe/H], on average (Figure \ref{fig:hist}(b)).  
In Figure \ref{fig:hist}(c), we show the [Fe/H]-distribution 
for different bins of galaxy age, where the peak is at lower $\text{[Fe/H]}$ for older stellar populations. 
Note that most of the ETGs in our sample have ages $7\le\tau_{\text{gal}}<9\,\text{Gyr}$ and that there are very few ETGs 
which have ages $6\le\tau_{\text{gal}}<7\,\text{Gyr}$ and $[\text{Fe/H}]\lesssim-1.0$.

\subsection{The age-dependence}

In Figure \ref{fig:ofe_feh}(a), we study the age-dependence of the $[\text{O/Fe}]$--$[\text{Fe/H}]$ relations in our sample of ETGs; 
different colours in the figure correspond to different average stellar ages of the galaxies. 
[O/Fe] ratios are, on average, almost constant at $[\text{Fe/H}]\lesssim-2$, independent of stellar age. 
Then, we find that ETGs with older 
ages have higher average $[\text{O/Fe}]$ at fixed $[\text{Fe/H}]$.
Finally, such a difference between chemical evolution tracks 
with different age steadily diminishes with increasing $[\text{Fe/H}]$. 

To understand the predicted behaviour of $[\text{O/Fe}]$ versus $[\text{Fe/H}]$, we recall that at very low $[\text{Fe/H}]$, 
the $[\text{O/Fe}]$ ratios mostly reflect the chemical enrichment from core-collapse SNe, which produce both O and Fe on a short timescale after the 
star formation event.
The IMF-weighted core-collapse SN yields determine a so-called ``plateau'' in the chemical evolution tracks. 
Then, as we move towards higher and higher $[\text{Fe/H}]$, 
the models predict a knee in the $[\text{O/Fe}]$--$[\text{Fe/H}]$ diagram, which is due to the larger amounts of Fe that SNe Ia 
inject in the galaxy ISM on a longer timescale. In Figure \ref{fig:ofe_feh}(a), the position of this knee in the $[\text{O/Fe}]$--$[\text{Fe/H}]$ diagram 
turns out to depend on the galaxy stellar age.  

ETGs with SFH peaked at very high redshift have the knee in the $[\text{O/Fe}]$--$[\text{Fe/H}]$ diagram at higher $[\text{Fe/H}]$ 
than young ETGs. 
This means that, in such oldest galaxies, when SNe Ia start to contribute, 
the $[\text{Fe/H}]$ of the ISM is already very high; this can only happen \textit{(i)} if 
the SFR in the oldest galaxies is more intense and more temporally concentrated than in the youngest ETGs, \textit{and} 
\textit{(ii)} if the chemical enrichment efficiency in the oldest ETGs is higher than in the youngest ones 
(or, equivalently, if the chemical enrichment timescales are shorter). 
Note that, although short star formation timescales can be inferred from the observed stellar ages or [$\alpha$/Fe] ratios, 
short chemical enrichment timescales can be clearly confirmed 
with the $[\alpha\text{/Fe}]$--$[\text{Fe/H}]$ relations. 
Even with rapid star formation, 
the evolutionary tracks of $[\alpha\text{/Fe}]$--$[\text{Fe/H}]$ 
cannot be described with 
the so-called  ``closed-box model'' (where the gas fraction is already high when the galaxy formed),  
but only with the so-called  ``infall model'', where the initial gas fraction is small and the star formation 
is directly driven by the gas infall \citep{tinsley1980,kobayashi2000}. 
The loss of metals by SN- and AGN-driven outflows does not appear at early stages in this figure, 
which is reasonable because of the deep potential well of the majority of galaxies and of the co-evolution of the AGN and its host galaxy. 


The difference in the chemical evolution tracks becomes smaller with increasing [Fe/H]. 
This is a signature of the fact that 
the chemical enrichment from SNe Ia dominates at high [Fe/H] 
over all the other nucleosynthetic processes, causing the ISM to saturate towards a common level of $[\alpha/\text{Fe}]$ ratios for all galaxies, despite their 
average stellar age.

Our final consideration on Figure \ref{fig:ofe_feh}(a) is about the large scatter in the predicted $[\text{O/Fe}]$ ratios 
at very low $[\text{Fe/H}]$, for ETGs with ages in the range $6\le \tau_{\text{gal}}<7\,\text{Gyr}$; 
this is due to the very low number of stellar populations with very low $[\text{Fe/H}]$ within this age bin (see also Figure \ref{fig:hist}c). 
[O/Fe] ratios of such stars reflect the variation in the yields of individual core-collapse SNe; more massive ($\sim 40\,\text{M}_{\sun}$) 
SNe give higher [O/Fe] than low-mass ($\sim 15\,\text{M}_{\sun}$) SNe \citep{kobayashi06}.
This effect is included in the inhomogeneous enrichment in our chemodynamical code  \citep[see][for more details]{kobayashi2011a}.

\begin{figure}
\centering
\includegraphics[scale=0.45]{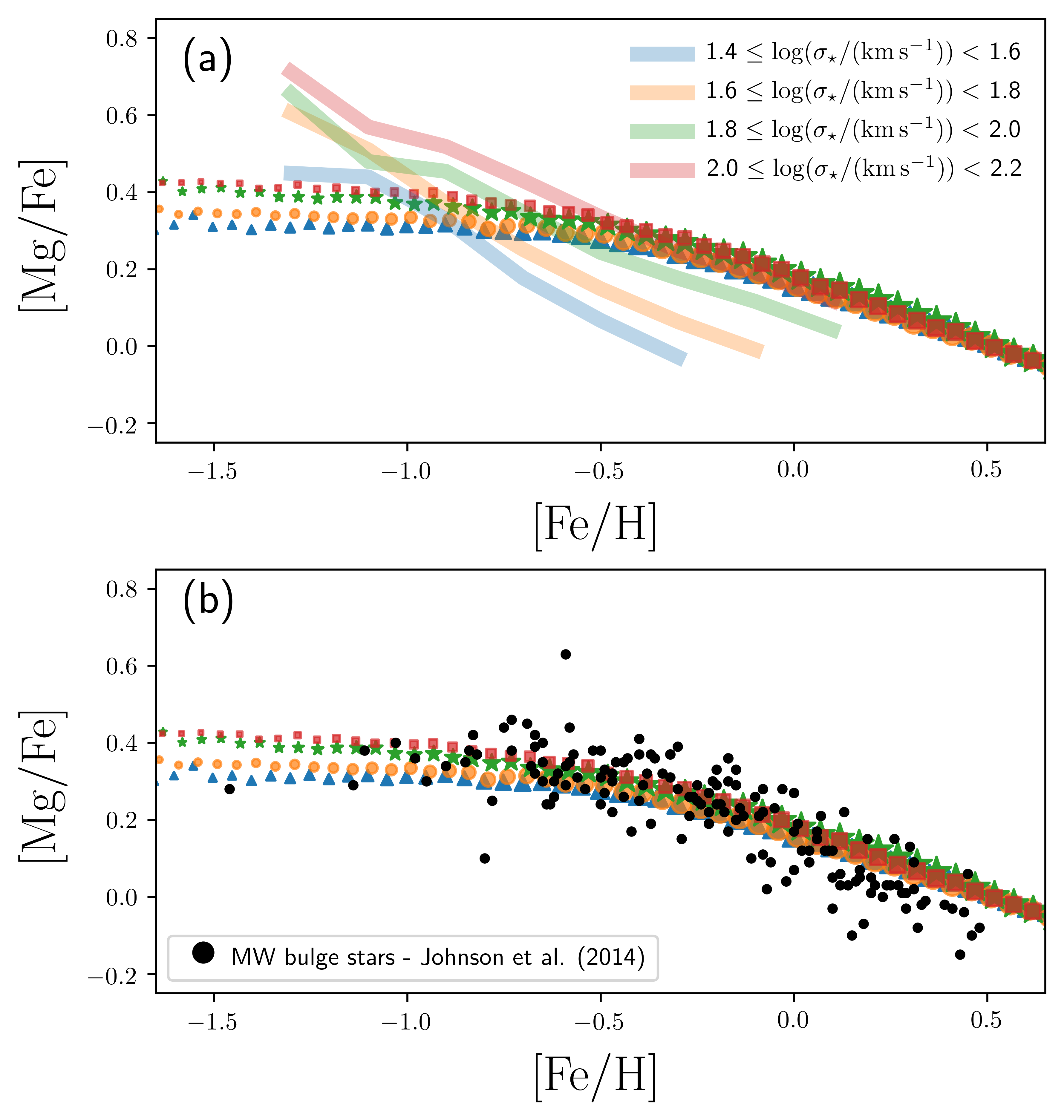}
\caption{The $[\text{Mg/Fe}]$--$[\text{Fe/H}]$ chemical evolution tracks. 
In panel (a), the solid curves represent the observed data of \citet{sybilska2018} 
and the points represent our model predictions, where different colours correspond to 
different stellar velocity dispersions of the galaxies. In panel (b), the black points represent the 
abundance measurements in the MW bulge stars from \citet{johnson2014}. }
\label{fig:mgfe_feh_sigma}
\end{figure}

\subsection{The mass-dependence}

In Figure \ref{fig:ofe_feh}(b), we study the mass-dependence of the $[\text{O/Fe}]$--$[\text{Fe/H}]$ relation for the same sample of galaxies 
as in Figure \ref{fig:ofe_feh}(a). 
In \citet[their figure 10]{taylor2015a}, which used the same simulation as in this Letter, it was shown that 
more massive ETGs are also older, on average. 
Nevertheless, the age-mass relation is not perfect especially for less-massive/young ETGs; for this reason, we find a different 
dependence of [O/Fe]--[Fe/H] as a function of the galaxy stellar mass. 
In particular, we predict [O/Fe] plateau values which are smaller for less massive ETGs, on average. 


The [O/Fe] plateau value depends on the relative contribution between core-collapse and SNe Ia, 
before SNe Ia become predominant. 
In less-massive ETGs, it took longer to reach $\text{[Fe/H]}=-1$ because of the lower SFRs, 
and young stars can form from low-metallicity ISM already enriched by SNe Ia; in massive ETGs, there is not so much low-metallicity ISM. 
In addition, a part of the early chemical enrichment from core-collapse SNe might be lost through SN-driven outflows in the shallower potential wells of less-massive systems. 
Finally, the variation in [O/Fe] at very low [Fe/H] is the consequence of an inhomogeneous chemical enrichment \citep{vincenzo2018}, 
which is particularly important for low-mass galaxies. 
This is one of the main features of chemodynamical simulations \citep{kobayashi2011a}, 
which cannot be obtained by classical one- or multi-zone chemical evolution models. 



\subsection{Comparing the model with observations} \label{subsec:comparison}

The observed data set is taken from \citet{sybilska2018}, which provide 
$[\text{Mg/Fe}]$, metallicity, and stellar velocity dispersions for a sample of $258$ ETGs in the 
ATLAS$^{\text{3D}}$ project \citep{cappellari2011}; we remark on the fact that most of these 
galaxies are spatially resolved, thanks to the SAURON integral field unit (IFU) instrument at the William Herschel Telescope; 
$58$ galaxies in the sample of \citet{sybilska2018} lie in the Virgo cluster, and the remaining $200$ galaxies lie in the field/group 
environment. 

It is important to note that the iron abundances in \citet{sybilska2018} have been estimated 
with the following empirical formula: $\text{[Fe/H]} = [Z/\text{H}]-0.75\times\text{[Mg/Fe]}$ \citep{vazdekis2015}, where 
$[Z/\text{H}]$ represents the abundance of all metals. 
This might cause an uncertainty in the derivation of chemical abundances.

In our simulation, we did not follow Mg, and we use O as a representative of $\alpha$-elements.
All $\alpha$-elements follow a very similar [$\alpha$/Fe]-[Fe/H] relations as O shows.
In the adopted nucleosynthesis yields, [O/Mg] is almost zero, independent of metallicity, which was consistent with 
local thermodynamic equilibrium (LTE) observations of stars in the solar neighbourhood \citep{kobayashi06}.
However, recent non-LTE observations suggested $\text{[O/Mg]} \sim 0.2$ at a wide range of metallicity \citep{zhao16}. 
Therefore, for the comparison to the galaxy observation, 
we apply a correction of $-0.2$ to the predicted $[\text{O/Fe}]$ ratios of our simulation. 

In Figure \ref{fig:mgfe_feh_sigma}(a), we compare our predicted $[\text{Mg/Fe}]$--$[\text{Fe/H}]$ relations (points with different colours) with 
the observations (solid curves);
in the figure, different colours correspond to different bins of stellar velocity dispersion, $\sigma$. 
The observations show parallel trends of $[\text{Mg/Fe}]$ against $[\text{Fe/H}]$,
where more massive ETGs have higher $[\text{Mg/Fe}]$ at fixed $[\text{Fe/H}]$. 
Our cosmological chemodynamical simulations reproduce the slope of [Mg/Fe] at $\text{[Fe/H]} \gtrsim -0.5$ of massive galaxies very well.
We also see a similar mass-dependence of 
the $[\text{Mg/Fe}]$--$[\text{Fe/H}]$ relations. However, our correlation 
is weaker than in the observations. 
At $\text{[Fe/H]} \lesssim -0.5$, our slope of $[\text{Mg/Fe}]$ versus $[\text{Fe/H}]$ is shallower 
than in the observations.
We also note that 
the majority of our massive ETGs are dominated by metal-rich stellar populations 
(Fig. \ref{fig:hist}), while most of the galaxies in the 
\citet{sybilska2018} sample show sub-solar metallicities. 
This is already seen in the predicted MZR, where our simulated galaxies 
are in good agreement with other observations, but show higher metallicities than in ATLAS$^{\text{3D}}$ \citep{taylor2016}.

The observed trend for the least massive ETGs could be reproduced, for example, by assuming stronger SN and/or AGN feedback associated with metal loss 
in less-massive systems. 
Alternatively, additional sub-classes of SNe Ia such as sub-Chandrasekhar SNe Ia or the so-called SNe Iax could produce a sharp [$\alpha$/Fe] decrease 
at lower $\text{[Fe/H]}$ \citep{kobayashi2015}; 
finally, the observed data might suggest a variable IMF, which -- in turn -- would affect our net yields of metals; 
in particular, a bottom-heavy IMF would give low [$\alpha$/Fe]. 
If the IMF is more bottom-heavy for low-mass galaxies, that might explain the observations (see also \citealt{kobayashi2010,demasi2018}), however, 
this is the opposite from what is suggested by other observations such as \citet{cenarro2003,parikh2018}. 

Our final consideration on Figure \ref{fig:mgfe_feh_sigma}(a) is that there might be an uncertainty in the calibration of observations; 
moreover, the absence of a plateau in the observed data might be due intrinsic uncertainties in the IFU observations, because 
low $[\text{Fe/H}]$ abundances are observed in the outer galaxy regions, where the surface brightness becomes fainter. 


In Figure \ref{fig:mgfe_feh_sigma}(b), the $[\text{Mg/Fe}]$--$[\text{Fe/H}]$ relations in our simulated ETGs are  compared with the observed 
chemical abundance measurements in the MW stellar bulge by \citet{johnson2014}. We find that the 
chemical evolution tracks in our simulated ETGs are qualitatively in agreement with the observations in the Galactic bulge. We notice that the decreasing trend of 
$[\text{Mg/Fe}]$--$[\text{Fe/H}]$ is more similar to the observations of \citet{sybilska2018} for ETGs with 
$1.8 \lesssim \log\big(\sigma_{\star}/(\text{km/s})\big) \lesssim 2.2$, which is consistent with the average stellar velocity 
dispersions in the Galactic bulge \citep{johnson2014}.

\section{Conclusions} \label{sec:conclusions}

We studied how the average $[\alpha\text{/Fe}]$--$[\text{Fe/H}]$ relations in a catalogue of simulated 
ETGs correlate with the galaxy stellar mass and age. 
We also compared the predictions of our cosmological chemodynamical simulations with AGN feedback with 
the latest observational results of \citet{sybilska2018}. Our main conclusions can be summarised as follows.

\begin{enumerate}

\item At low [Fe/H], we predict an $[\alpha/\mathrm{Fe}]$ plateau, with higher values for more massive galaxies. 
The position of the knee in [$\alpha$/Fe]--[Fe/H] is determined both by the galaxy SFH and by the chemical enrichment timescale, 
and it is more sensitive to the average galaxy stellar age than mass. 

\item Our simulations can qualitatively explain the observed mass-dependence of $[\text{Mg/Fe}]$--$[\text{Fe/H}]$, 
but quantitatively there are mismatches. 
In particular, less massive ETGs in the observations show even earlier decrease in $\text{[Mg/Fe]}$ than in our model.
This suggests even lower chemical enrichment efficiencies in low-mass ETGs than in our model. 

\item There might be an uncertainty in the calibration of observations, our net yields might be too high, 
stronger AGN/SN feedback or additional classes of SNe Ia might be required. 
Also an IMF which is more bottom-heavy for less-massive systems might help to reproduce the observational data.



\end{enumerate}

%



\section*{Acknowledgments}
\begin{small}
We thank an anonymous referee for many comments and suggestions. 
We thank V. Springel for providing \textsc{Gadget-3}, A. Sybilska and H. Kuntschner for providing the observational data. 
FV and CK acknowledge funding from STFC (ST/M000958/1). 
PT acknowledges funding through a Discovery Projects grant from the Australian Research Council (grant no. DP150104329). 
This work used the DiRAC Data Centric system (part of the National E-Infrastructure), funded by Durham University and grants ST/K00042X/1, ST/K00087X/1 \& ST/K003267/1. 
\end{small}

\end{document}